# Online classification of imagined speech using functional near-infrared spectroscopy signals

Alborz Rezazadeh Sereshkeh[1,2], Rozhin Yousefi[1,2], Andrew T Wong[1,2], Tom Chau[1,2]

[1] Institute of Biomaterials and Biomedical Engineering, University of Toronto, Canada
[2] Bloorview Research Institute, Holland Bloorview Kids Rehabilitation Hospital, Toronto, Canada

E-mail: tom.chau@utoronto.ca

## Abstract

Most brain-computer interfaces (BCIs) based on functional near-infrared spectroscopy (fNIRS) require that users perform mental tasks such as motor imagery, mental arithmetic, or music imagery to convey a message or to answer simple yes or no questions. These cognitive tasks usually have no direct association with the communicative intent, which makes them difficult for users to perform. In this paper, a 3-class intuitive BCI is presented which enables users to directly answer yes or no questions by covertly rehearsing the word "yes" or "no" for 15 s. The BCI also admits an equivalent duration of unconstrained rest which constitutes the third discernable task. Twelve participants each completed one offline block and six online blocks over the course of 2 sessions. The mean value of the change in oxygenated hemoglobin concentration during a trial was calculated for each channel and used to train a regularized linear discriminant analysis (RLDA) classifier. By the final online block, 9 out of 12 participants were performing above chance ($p<0.001$), with a 3-class accuracy of 83.8±9.4%. Even when considering all participants, the average online 3-class accuracy over the last 3 blocks was 64.1±20.6%, with only 3 participants scoring below chance ($p<0.001$). For most participants, channels in the left temporal and temporoparietal cortex provided the most discriminative information. To our knowledge, this is the first report of an online fNIRS 3-class imagined speech BCI. Our findings suggest that imagined speech can be used as a reliable activation task for selected users for development of more intuitive BCIs for communication.

Keywords: brain-computer interfaces, functional near-infrared spectroscopy, imagined speech, regularized linear discriminant analysis

## 1. Introduction

Brain-computer interfaces (BCIs) can be used to provide a communication channel for individuals with severe motor impairments who are unable to communicate independently [1]. Since the emergence of BCIs, various activation protocols have been suggested and tested. A subset of these protocols are known as reactive BCIs [2], which require the user to attend to external stimuli. Examples include P300 spellers [3] and BCIs based on steady-state visually evoked potentials [4]. BCI protocols that do not require an external stimulus give rise to active BCIs [2], where instead, users perform a mental task. Some of the most common examples of these mental tasks are motor imagery [5], mental arithmetic [6] and word generation [7]. Given an adequate classification accuracy, a BCI user can perform each of these mental tasks to convey a different message, e.g. to answer yes or no questions. However, these mental tasks are usually difficult to perform by the target population since the tasks are non-intuitive and unrelated to the actual intended message.

An intuitive mental task for BCIs which has attracted attention during the last decade is imagined speech - also known as covert speech [8]. A review of reported BCIs based on imagined speech and their performances are provided in [8] and [9]. According to these reviews, invasive measurement techniques such as electrocorticography (ECoG) have been required in most cases where accuracies of classifying electrophysiological brain signals during imagined speech have exceed 70% (the touted threshold for practical BCI application [10]) [11-13]. In contrast, most BCIs based on non-invasive electrophysiological measurements, including electroencephalography (EEG) and magnetoencephalography (MEG), have yielded accuracies less than 70% when discriminating between two different imagined speech tasks [14-16]. Moreover, only one study used a real-time paradigm which reported an average classification accuracy of ~69% using EEG signals recorded during covert repetition of "yes" and "no" [17].

Another brainwave response which has been investigated during speech related tasks is the hemodynamic response [18]. Initial studies on the hemodynamic response related to speech generation and comprehension deployed positron emission tomography (PET) and functional magnetic resonance imaging (fMRI) to study activated brain areas [19]. A review of these studies is provided in [19].

Initial studies to investigate the reliability of the hemodynamic response to decode speech focused on the averaged hemodynamic response over many repetitions of a speech task [8]. However, a successful imagined speech BCI should be able to decode speech in a single trial [8]. Several studies used fMRI to discriminate between brain patterns activated when different nouns [20] and Dutch vowels [21] were presented either aurally or visually to participants. In [22], covert repetition of a nursery rhyme was used as an activation task (along with mental calculation and two motor imagery tasks) in a 4-class BCI based on fMRI, and yielded an average classification accuracy greater than 90%. However, due to the limitations of fMRI, the duration of each trial was relatively long (~2 min). More importantly, fMRI cannot be used in development of a portable BCI.

Another modality to measure the hemodynamic response is functional near-infrared spectroscopy (fNIRS). An fNIRS device can be portable, and the duration of each trial can be as short as 10-15 seconds [23]. Early applications of fNIRS in speech recognition focused on distinguishing among different speech modes: overt, silent and imagined speech, and trials without any speech activity [24-25]. In [24], each speech task included a whole sentence, and different speech modes were successfully discriminated using fNIRS data. In another fNIRS study, different patterns of hemodynamic responses were reported during trials of inner recitation of hexameter or prose, with mental arithmetic as a control task [26].

Due to the slow nature of the hemodynamic response, decoding small units of language, such as nouns, is more difficult compared to full sentences or different speech modes [8]. Gallegos-Ayala et al. reported an fNIRS-BCI for answering "yes" or "no" questions. This BCI was tested on a patient with amyotrophic lateral sclerosis (ALS) who answered different questions by simply thinking "yes" or "no" [27]. The duration of each trial was 25 s and an online classification accuracy of 71.7% was reached for this patient.

Hwang et al. tested a similar "yes" or "no" paradigm on eight able-bodied participants using fNIRS [28]. The duration of each trial was reduced to 10 s. Different types of hemodynamic features, feature numbers and time window sizes were tested and their accuracies were compared. An offline average accuracy of ~75% was reported when the best feature set was employed for each participant. They also reported that kurtosis features yielded the highest average classification accuracy among different types of features. Surprisingly, the location of the fNIRS channels did not cover any of the temporal regions which are some of the most important speech-related brain areas.

In [29], Chaudhary et al. expanded the work presented in [27]. Four ALS patients used the same fNIRS-BCI to answer yes or no questions by thinking "yes" or "no". Three participants completed more than 46 sessions, each containing 20 questions, and one participant completed 20 sessions. An average online classification accuracy of more than 70% (above the chance-level) was reported across participants.

As summarized, none of the previous online non-invasive, portable neuroimaging studies (EEG and fNIRS) have investigated classification of more than 2 classes. The classification was either limited to imagined speech versus a control condition (e.g. rest) or between two imagined speech tasks.

In this study, we developed an fNIRS-BCI for online 3-class classification of the following three tasks: thinking "yes" while mentally rehearsing the phrase "yes", thinking "no" while mentally rehearsing the phrase "no", and unconditional rest. To the best of our knowledge, this is the first 3-class BCI based on imagined speech using a portable and non-invasive neuroimaging technique, i.e. EEG or fNIRS.

## 2. Methods

### 2.1 Participants

Twelve able-bodied participants (7 males) between the ages of 23 and 33 (mean age: 28.4±2.9 years) participated in this study. Participants were fluent in English, had normal or corrected-to-normal vision, and had no health issues that could adversely affect the measurements or their ability to follow the experimental protocol. These issues included neurological, cardiovascular, respiratory, psychiatric, metabolic, degenerative, or alcohol-related conditions. Participants were asked to refrain from drinking alcoholic or caffeinated beverages at least 3 hours prior to each session. This study was approved by the research ethics boards of the Holland Bloorview Kids Rehabilitation Hospital and the University of Toronto. Written consent was obtained from all participants prior to study participation.

## 2.2 Instrumentation

NIRS measurements were collected from the frontal, parietal and temporal cortices using a continuous-wave near-infrared spectrometer (ETG-4000 Optical Topography System, Hitachi Medical Co., Japan). As shown in figure 1, 16 NIR emitters and 14 photodetectors were integrated in two 3 × 5 rectangular grids of optical fibers in a standard EEG cap (EasyCap, Germany). Each NIR emitter contained two laser diodes that simultaneously emitted NIR light at wavelengths of 695 nm and 830 nm. The optical signals were sampled at 10 Hz.

Adjacent positions in each of the two 3 × 5 grids, were 3 cm apart. Only optical signals arising from source-detector pairs (or 'channels') separated by 3 cm were acquired for analysis. This separation distance yielded a depth penetration of light between 2 and 3 cm [30, 31], which surpasses the average scalp-to-cortex depth within the brain areas monitored [32]. Using this configuration, optical signals were acquired from a total of 44 measurement sites on the cerebral cortex, 22 on each hemisphere (see figure 1). In addition to NIRS measurements, EEG signals were recorded from 32 locations using BrainAmp DC amplifier (Brain Products GmbH, Germany). These data are not analyzed herein.

## 2.3 Experimental protocol

Participants attended two sessions on two separate days. The first session consisted of three blocks, starting with an offline block and followed by two online blocks. In the offline block, participants performed 36 trials, including 12 "yes" imagined speech trials, 12 "no" imagined speech trials and 12 unconstrained rest trials. The trials were presented in a pseudorandom order. At the end of the offline block, a 3-class classifier was trained using the data from the offline block. Each online block consisted of 24 trials, 8 trials per class, presented in a pseudorandom order. Participants were presented with the classifier decision subsequent to each trial. The 3-class classifier was re-trained after each block using the data from all previous blocks.

The second session consisted of four online blocks, each with 24 trials equally distributed among the three classes presented in pseudorandom order. Similar to the first session, the 3-class classifier was retrained after each block. The timing diagram is depicted in figure 2.

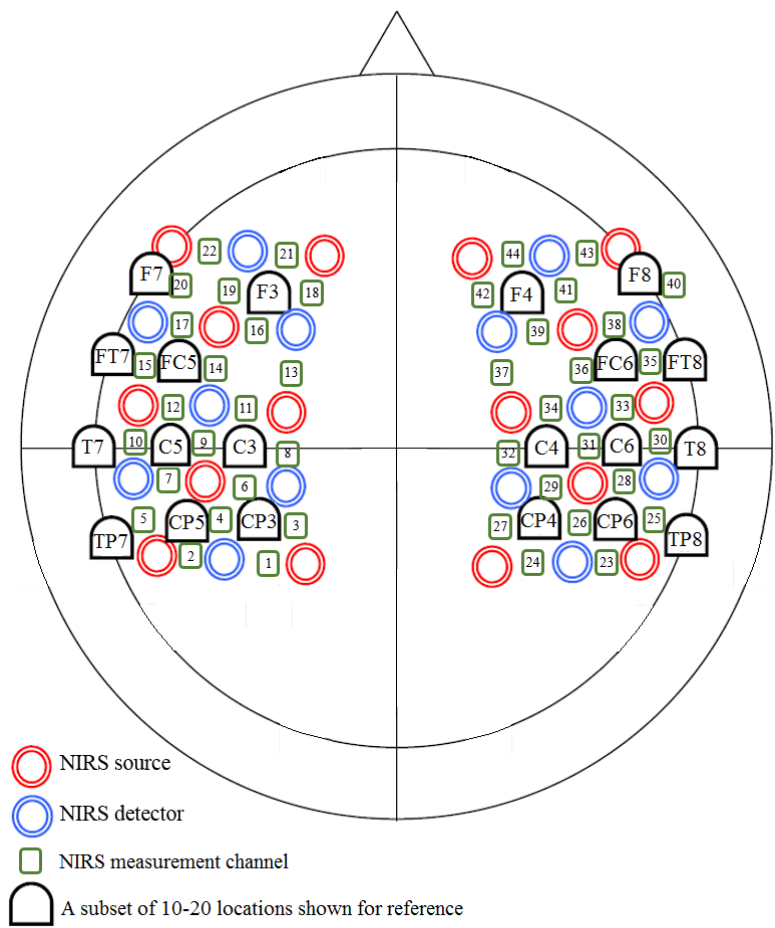

**Figure 1.** The placement of NIRS sources and detectors. A subset of 10-20 locations are also shown for reference.

A fixation cross appeared at the center of a blank screen at the beginning of each trial and persisted throughout the trial. Each trial started with a 14 s baseline period which allowed the hemodynamic signal to return to a basal level [33]. Participants were asked to refrain from performing any of the imagined speech tasks during this period. They had no knowledge of the type of the next trial at the time of baseline collection.

In the imagined speech trials, a question appeared on the screen after the baseline period for 3 s. Then it was replaced by the instruction "start", which disappeared after 1 s. The question was always the same: "Is this word in uppercase letters? WORD". For the yes trials, the word was written in uppercase letters. For the no trials, the word was written in lowercase letters. The words were different in each question and were selected at random from a list of emotionally neutral words suggested by [34]. In the unconstrained rest trials, the phrase "rest" appeared on the screen for 3 s, which was then replaced by the instruction, "start", for 1 s.

Participants were instructed to commence the mental task as soon as the "start" instruction disappeared. For the imagined speech trials, participants were instructed to think "yes" or "no" while iteratively repeating the word "yes" or "no" mentally. They were explicitly instructed to perform the task without any vocalization or motor movement, especially of the lips, tongue or jaw. In the unconstrained "rest" trials, participants allowed normal thought processes to occur without restriction. The participant was asked to perform the mental task for 15 s for all trial types. This duration was determined based on previous similar fNIRS studies and the suggested minimum measurement time for a hemodynamic response in literature [23].

At the end of each session, the participants were asked to rate from 1 to 5 (where 1 was the lowest and 5 the highest) their perceived ability to perform the task (data not shown).

## 2.4 Data analysis

### 2.4.1 Signal processing.

First, we converted optical intensities to oxygenated hemoglobin concentration changes, denoted as [HbO], using the modified Beer-Lambert law [35]. The signals were then filtered using a using a $3^{rd}$ order Chebyshev type II low-pass filter with a passband cutoff frequency of 0.1 Hz, passband ripple of 0.1 dB, stopband cut off frequency of 0.5 Hz and minimum stopband attenuation of

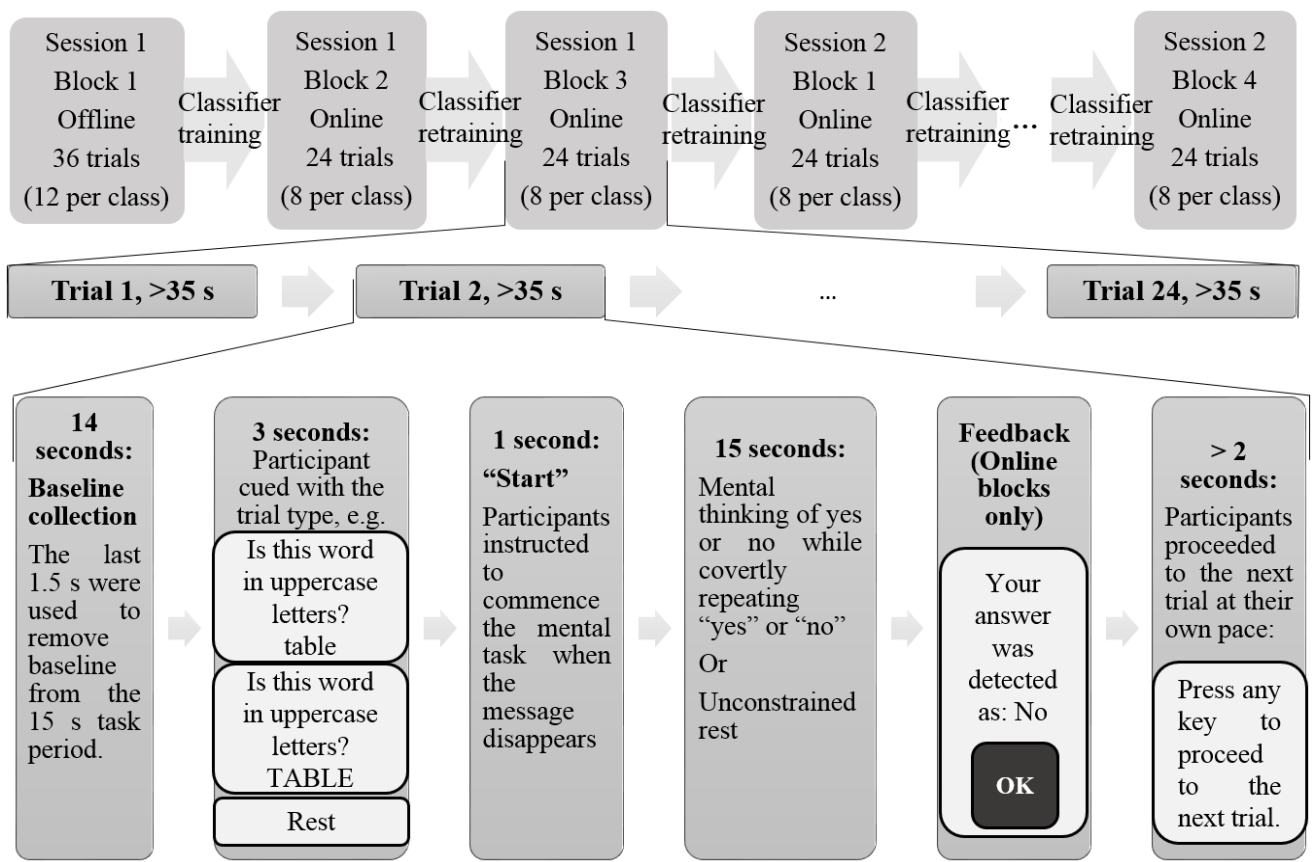

**Figure 2.** The timing diagram of the experiment

50 dB. This filter removed any high frequency physiological noise, including Mayer waves at 0.1 Hz, respiration at ~0.3 Hz and cardiac activity at 0.8-1.2 Hz.

*2.4.2 Baseline removal.* Fluctuations in the value of HbO are not limited to the periods of various cognitive tasks. The baseline value of HbO can change from one day to another or even from the beginning to the end of a session [36]. Hence, some BCI studies have added baseline collection periods to the beginning of each session or block to adjust for this natural fluctuation [36-37].

In this study, baseline data were collected prior to each trial to calculate a more precise and trial-specific mean baseline value. From the 14 s baseline period, we calculated the mean of [HbO] during the last 1500 ms for each fNIRS channel and subtracted this value from the subsequent trial on a per-channel basis. The last 1.5 s was chosen instead of the entire 14 s since the hemodynamic signal requires approximately 12 s to return to its baseline value after any cognitive load [33].

*2.4.3 Feature extraction.* The mean value of the oxygenated hemoglobin concentration change for each channel during the entire length of each trial were used as features for classification. Hence, each trial was represented as a 1×44 vector of features (44 channels x 1 feature).

Other common types of NIRS features are variance, slope, skewness and kurtosis of changes in oxygenated, deoxygenated, and total hemoglobin concentrations. These features were examined during pilot sessions, but the mean of [HbO] led to the highest classification accuracy and therefore was selected to provide real-time feedback during the online trials. This feature has been previously used in a similar "yes" vs "no" fNIRS study on ALS patients [29]. Furthermore, it has been shown in another "yes" vs "no" study on healthy participants [28] that features extracted from oxygenated hemoglobin concentrations provide more discriminative information than features derived from deoxyhemoglobin concentrations.

*2.4.4 Classification.* For classification, a regularized linear discriminant analysis (RLDA) algorithm was used [38]. This method was chosen as it led to the highest average accuracy during the pilot sessions compared to support vector machines (linear, polynomial, radial basis function and sigmoid kernels), neural networks (multilayer perceptron with one hidden layer) and naïve Bayes classifiers.

To discriminate between the 3 classes, a multiclass LDA was used for classification. In contrast with other types of discriminant analysis, e.g. quadratic discriminant analysis, LDA assumes that all classes have the same covariance. This common pooled covariance matrix is defined as:

$$\hat{\Sigma} = \sum_{k=1}^{K} \sum_{i \in I_k} (X_i - \mu_k)(X_i - \mu_k)^T / (N - K) \quad (1)$$

where K is the number of classes, $X_i$ is the feature vector for the $i^{th}$ example, $I_k = \{i \mid y_i = k\}$ is the subset of indices identifying the examples of the $k^{th}$ class, $y_i$ is the class label of the $i^{th}$ example, $\mu_k$ is the mean of all examples of the $k^{th}$ class, and N is the total number of examples.

LDA classification is done based on the analysis of the following two scatter matrices: the within-class scatter matrix and the between-class scatter matrix. The within-class scatter matrix can be expressed in terms of the common covariance matrix defined in equation (1):

$$S_w = (N - K) \times \hat{\Sigma} \quad (2)$$

The between-class scatter matrix is defined as:

$$S_b = \sum_{k=1}^{K} N_k (\mu_k - \mu)(\mu_k - \mu)^T \quad (3)$$

where $\mu$ is the overall mean of all examples and $N_k$ is the number of examples in the $k^{th}$ class, or $N_k = |I_k|$ where $|\cdot|$ denotes cardinality.

The main goal of LDA is to find a set of coefficients, W, that maximizes the following ratio:

$$W_{LDA} = \underset{W}{\arg\max} \frac{W^T S_b W}{W^T S_w W} \quad (4)$$

This ratio is called the Fisher criterion.

In regularized LDA, the common pooled covariance matrix is replaced with the following covariance matrix for each class:

$$\hat{\Sigma}_k(\gamma) = (1 - \gamma)\hat{\Sigma} + \gamma \hat{\Sigma}_k \quad (5)$$

where $\hat{\Sigma}$ is the common pooled covariance matrix defined in equation (1), $\gamma$ is the regularization parameter, and $\hat{\Sigma}_k$ is the covariance matrix of the $k^{th}$ class, defined as:

$$\hat{\Sigma}_k = \frac{1}{N_k - 1} \sum_{i}^{y_i=k} (X_i - \mu_k)(X_i - \mu_k)^T \quad (6)$$

It can be seen that when $\gamma$ is equal to zero, $\hat{\Sigma}_k(\gamma)$ is equal to $\hat{\Sigma}$, and the optimization equation will be the same as a non-regularized LDA.

*2.4.5 Optimization of the regularization parameter.* The only hyper-parameter in an RLDA classifier is the regularization parameter, gamma, which can be any value between 0 and 1. This parameter was optimized every time the classifier was trained using a leave-one-out cross-validation (LOOCV) method. Prior to each online block, we calculated the LOOCV accuracy on the data from all previous blocks for different gamma values in the range of 0 to 1 in 0.05 increments, with two exceptions which are explained below. The gamma which resulted in the highest LOOCV accuracy was selected for subsequent classifier training. In case of a tie, the largest gamma was selected to obtain a more generalized classifier. The classifier was then trained on the entire training set using that gamma.

Two restrictions were applied to the gamma range. Firstly, during each session, the maximum value of gamma in the gamma-optimization step was set to the gamma used in the previous block. Since more same-day data was acquired as the session progressed, the need for generalizing the classifier was reduced and the classifier could be further optimized. Secondly, at the beginning of the second session, a minimum of 0.3 was used for gamma to prevent overfitting, i.e. over-emphasis on data from the first session and thereby preserve generalizability. The data analysis steps are summarized in figure 3.

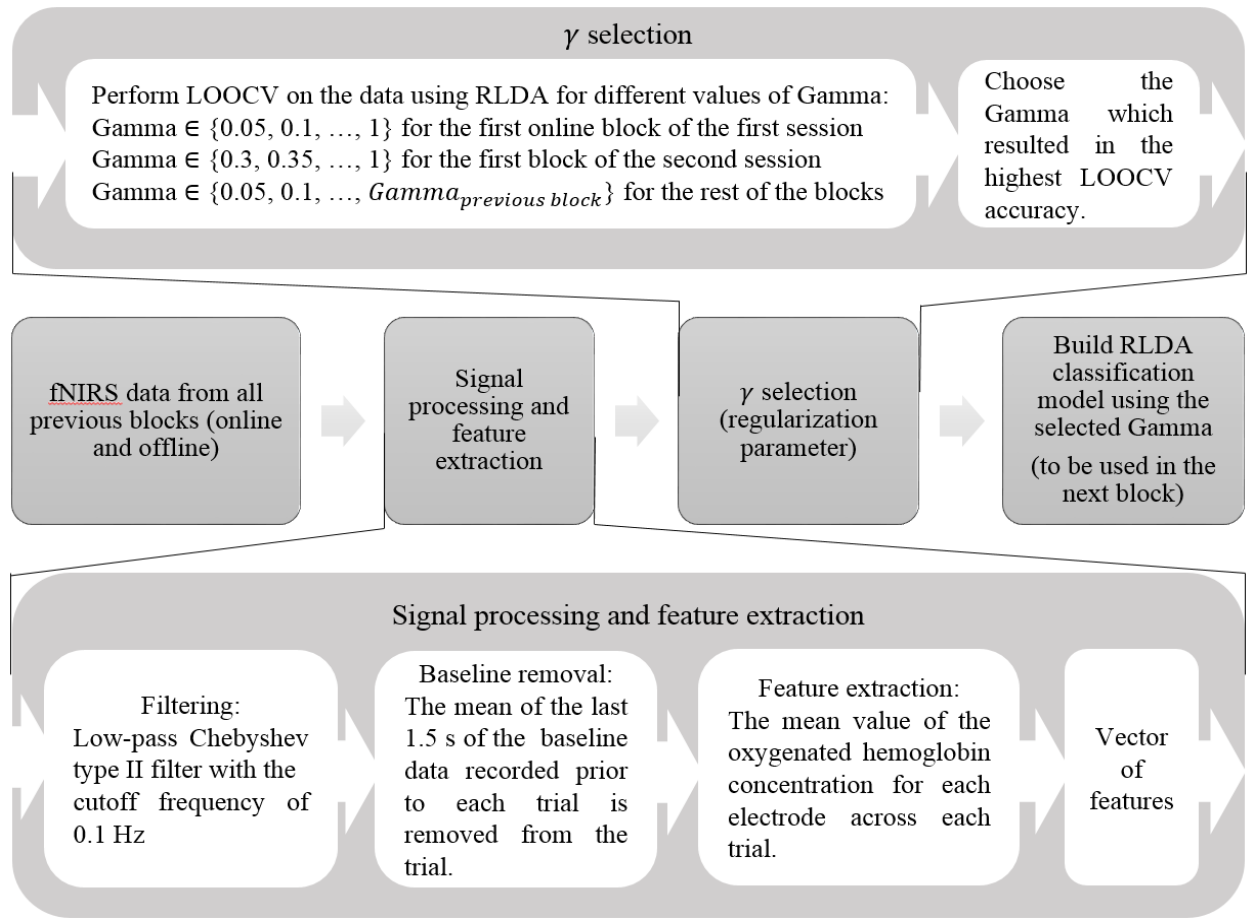

**Figure 3.** The mathematical steps for building a classifier prior to each online block (LOOCV = leave one out cross-validation, RLDA= regularized linear discriminant analysis)

## 3. Results

### 3.1 Online 3-class accuracies

Table I provides the online 3-class classification accuracies obtained during the six online blocks performed by each participant. Nine out of twelve participants reached above-chance online classification accuracy in their final three blocks (p<0.001), achieving an average online accuracy above the 70% threshold (minimum acceptable threshold for practical BCI applications [10]).

For participants P5, P7 and P11, the second session was interrupted as these participants asked to have the cap removed due to discomfort. After the removal of the cap, these participants took a short break and continued the experiment. In the post-session questionnaire, all three of these participants stated that the task was difficult (5 on the scale of 1 to 5) to perform given the discomfort of the cap. Hence, in Table 1, the mean accuracy is also reported without these 3 participants.

As seen in Table 1, the last online block yielded the highest average accuracy across participants, which was significantly higher than the first online block ($p = 0.022$ using Wilcoxon signed rank test). This increase in average accuracy is likely the combined effect of two factors: improved classifier robustness due to the accumulation of training data and more consistent task performance (and hence brain signals) by the user upon receiving real-time feedback [1].

At the beginning of the second session, there was a drop in the average accuracy. As the classifier was trained using data from a different day, this decline in accuracy may be attributable to slight variations in fNIRS cap positioning, changes in mental states between sessions (e.g. fatigue or attention [39]), or variations in metabolic states [40].

Table 1. Online 3-class accuracies (%) for each participant for all online blocks. Average accuracies exceeding the upper limit of the 95%, 99% and 99.9% confidence interval of chance are marked with *, ** and ***, respectively.

| Participant | Session 1 – Block 2 | Session 1 – Block 3 | Session 2 – Block 1 | Session 2 – Block 2 | Session 2 – Block 3 | Session 2 – Block 4 | Average of last 3 blocks |
|---|---|---|---|---|---|---|---|
| P1 | 50.0* | 87.5*** | 75.0*** | 75.0*** | 62.5*** | 79.2*** | 72.2*** |
| P2 | 37.5 | 45.8 | 37.5 | 83.3*** | 83.3*** | 95.8*** | 87.5*** |
| P3 | 75.0*** | 66.7*** | 58.3** | 66.7*** | 75.0*** | 79.2*** | 73.6*** |
| P4 | 75.0*** | 66.7*** | 41.7 | 66.7*** | 70.8*** | 83.3*** | 73.6*** |
| P5 | 41.7 | 54.2* | 37.5 | 33.3 | 45.8 | 25.0 | 34.7 |
| P6 | 95.8*** | 100*** | 87.5*** | 83.3*** | 100*** | 100*** | 94.4*** |
| P7 | 62.5*** | 58.3** | 62.5*** | 37.5 | 45.8 | 75.0*** | 52.8*** |
| P8 | 62.5*** | 41.7 | 41.7 | 58.3** | 50.0* | 83.3*** | 63.9*** |
| P9 | 45.8 | 50.0* | 41.7 | 54.2* | 50.0* | 70.8*** | 58.3*** |
| P10 | 83.3*** | 79.2*** | 62.5*** | 70.8*** | 66.7*** | 87.5*** | 75.0*** |
| P11 | 29.2 | 33.3 | 41.7 | 45.8 | 45.8 | 25.0 | 38.9 |
| P12 | 37.5 | 58.3** | 58.3** | 33.3 | 45.8 | 54.2* | 44.4* |
| Mean (all participants) | 58.0±21.0 | 61.8±19.4 | 53.8±16.2 | 59.0±18.3 | 61.8±17.8 | 71.5±24.7 | 64.1±20.6 |
| Mean (P1-P4, P6, P8-P10, P12)[1] | 62.5±21.0 | 66.2±19.7 | 56.0±17.2 | 65.7±15.7 | 67.1±17.6 | 81.5±13.5 | 71.5±16.7 |

[1]Excluding three participants, P5, P7 and P11, whose second sessions were interrupted upon participants' request to remove the cap due to discomfort and fatigue.

### 3.2 The role of different fNIRS channels in providing discriminative information

In order to determine the role of each fNIRS channels in providing the discriminative information, we used the value of the Fisher criterion calculated for each feature. Since RLDA was used for classification (which works based on maximizing the Fisher criterion) and each fNIRS channel produced only one feature, the calculated Fisher criterion for that feature represented the level of discriminative information that fNIRS channel provided. Figure 4(a) depicts the brain map of the calculated Fisher criterion for each channel averaged across participants. Fig 4(b) provides the brain map of the standard deviation of the calculated Fisher criterion across participants.

### 3.3 The role of regularization

Regularization can be necessary in classification models to preserve generalizability, especially when the number of samples are of the same order of magnitude as the number of features [41]. To determine whether using a regularization parameter was helpful for online classification, we retrospectively calculated the accuracies for all online blocks without any regularization. As evident in figure 5, regularization improved the average accuracies across all blocks across participants ($p = 1.3 \times 10^{-5}$ Wilcoxon signed rank test). However, when the classification accuracies with and without regularization were compared separately for each block, only the first three online blocks exhibited a significant difference ($p = 0.003, 0.016$ and $0.045$ for the first three blocks, respectively, using Wilcoxon signed rank test). The significant difference in early blocks confirmed the importance of regularization when the training dataset is relatively small, as well as when a BCI is trained only on the data from a previous day. The difference was not significant in the remaining blocks, the last three online blocks, due to the inclusion of more same-day data in the classifier training. The selected gamma values for all participants in different blocks as well as the changes across blocks can be found in figure 6.

## 4. Discussion

### 4.1 Comparison with previous multiclass fNIRS-BCIs and previous imagined speech BCIs

In this paper, we proposed an online 3-class BCI based on imagined speech. An average ternary classification accuracy of 71.5±24.7% was reached across all participants in their last block, with 9 out of 12 participants surpassing the chance level (p<0.001).

So far, only few studies have explored the possibility of developing a multiclass (>2 classes) BCI using fNIRS [42]. Power et al. [6] developed an fNIRS-BCI to classify between mental singing, mental arithmetic and unconstrained rest and reported an offline ternary classification accuracy of 56.2±8.7% across 7 participants. Herff et al. [43] used fNIRS to classify between three levels of the n-back task (where participants were instructed to continuously remember the last n letters of a series of rapidly flashing letters) and the rest state, reporting an average offline accuracy of 44.5±10.0% across 10 participants. Weyand et al [44] investigated different combinations of six cognitive tasks and reported an offline ternary accuracy of 60.5±6.0% across 10 participants. Recently, Schudlo et al [42] reported one of the first online 3-class fNIRS-BCIs. The three tasks included verbal fluency, Stroop task and rest and were differentiated online with an accuracy of 74.2±14.8% across 11 participants.

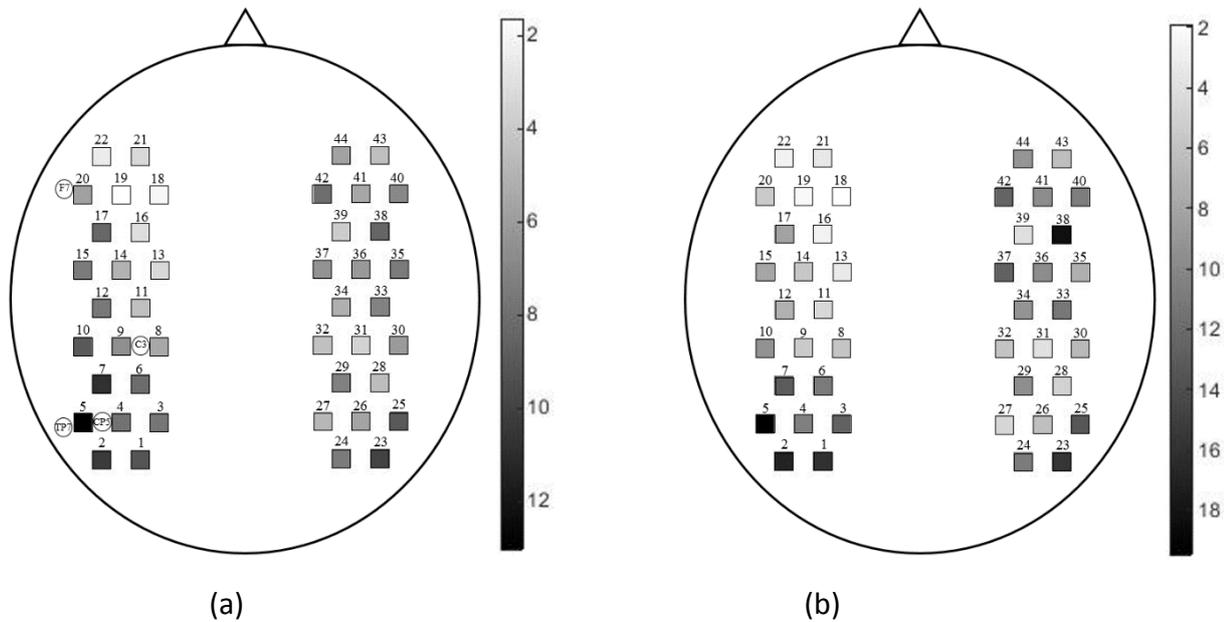

(a)                                                                                           (b)

**Figure 4.** The brain map of (a) the average of the Fisher criterion value across participants and (b) the standard deviation of the Fisher criterion value across participants.

Using a BCI for online classification of brainwaves when participants mentally think yes or no has been limited to two studies, one with EEG [17] and one with fNIRS [29]. Both studies reported ~70% average binary classification

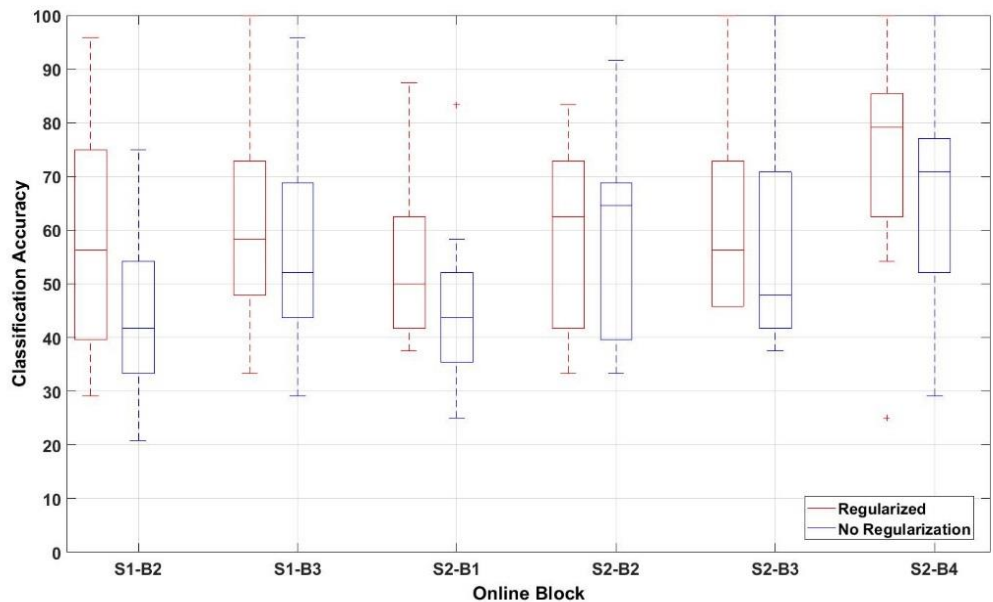

**Figure 5.** The classification accuracies in different online blocks averaged over all participants with (red) and without (blue) regularization. The notation Sn - Bm identifies the $m^{th}$ block (B) of the $n^{th}$ session (S).

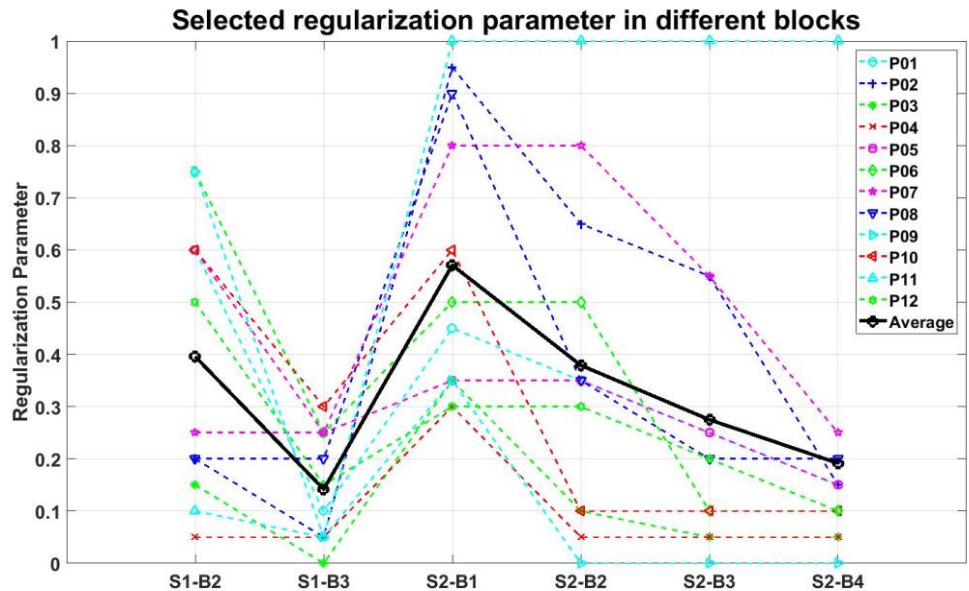

**Figure 6.** Changes in the selected regularization parameter, $\gamma$, for different participants across different blocks. The selected $\gamma$ is the value which provided the highest leave-one-out cross validation accuracy on the data from all previous blocks. The notation S n - B m identifies the $m^{th}$ block (B) of the $n^{th}$ session (S).

accuracies (see section I for a more extensive summary of these BCIs).

In terms of the average classification accuracy across participants, our results surpassed the outcome of all previous mentioned BCIs except [42]. However, the Stroop task used in [42] required users to attend to a screen which is not practical for individuals with visual impairments.

Our results exhibited higher standard deviation across participants, namely, 18.3%, 17.8% and 24.7% in the last three online blocks, compared to previously mentioned BCIs. The standard deviation increased in the last block since most participants obtained higher accuracies (e.g. 100% for the last online block of P6) as the session progressed while a few hovered at chance level accuracies across all blocks (e.g. 25% in the last online block for P5 and P11). If we were to exclude the participants who were unable to complete the session without interruptions, the accuracy in the last block would jump to 81.5±13.5%. This new standard deviation is in the same range as those previously reported for multiclass BCIs. Note that P5 and P11 had difficulties obtaining accuracies above chance in both sessions, and P7 experienced difficulties at the beginning of session 2. This is not unusual as certain individuals may have difficulties performing certain tasks or using certain BCI modalities.

*4.2 The role of different brain regions*

In Fig 4(a), we see that the channels in the left temporal and left temporoparietal regions yielded the highest Fisher criterion value, and therefore provided the most discriminative information. Channels 5, 7 and 2 provided the three highest average Fisher criterion values (13.02, 10.80 and 10.33, respectively). Channel 5 is located between CP5 and TP7, while channel 7 is positioned between CP5 and C5, and channel 2 is situated close to CP5. Although the exact Brodmann areas of these channels cannot be determined without an fMRI scan, previous concurrent EEG-fMRI studies can provide an estimation of the associated Brodmann regions of these channels. Based on the channel maps from [45], these three channels (5, 7 and 2) approximately cover parts of Brodmann areas 21, 22, 39, 40 and 42. These Brodmann areas represent in part, Wernicke's area, the left angular gyrus, the left supramarginal gyrus and the left auditory cortex. All these areas are belong to the speech network of the brain [46] and have been previously identified in other imagined speech studies as yielding discriminative information [19].

Another channel of note is channel 20, as it is close to F7 and Broca's area (see figure 1) [45]. Although Broca's area is known to play an important role in speech production, the average Fisher score of this channel was 5.80, ranking it 20th out of 44 channels. This finding is in line with several previous studies on the classification of imagined speech, where greater discriminative information was found in the temporal and temporoparietal regions close to Wernicke's area, compared to Broca's area, especially when the imagined speech task did not involve the production of complicated phrases [17, 19].

Fig 4(b) depicts the brain map of the standard deviation of the calculated Fisher criterion across all participants. Again, channel 5 provided the highest standard deviation of the Fisher criterion. The value of the Fisher criterion varied from 0.31 (P12) to 64.33 (P6) for this channel. The large variations in the Fisher scores of the speech-related regions may be attributable to inter-individual performance variations of the imagined speech task. As mentioned, all participants were instructed to think yes or no while covertly repeating the phrase without any motor movements. Since they were given the freedom to "think" yes or no in their own way, some individuals may have focused on the meaning of the phrases (affirmative versus negative response), the articulation of the phrases (with or without motor imagery of the articulation), or on imagining hearing the phrases, while covertly repeating them. For example, the highest Fisher criterion for participant 12 was obtained on channel 8 (located approximately between C3 and C1) with the value of 7.09, while channel 5 produced the lowest Fisher criterion at 0.31. The area covered by channel 8 is known to be activated during motor imagery tasks, which may indicate that this participant mainly focused on imagining motor movements required for speech production. The low accuracy of P12 compared to other participants seems to support the hypothesis that P12 may have utilized a unique approach to covert speech.

The variation across participants in the location of the channels which provided the maximum discriminative information for classification has been frequently reported in previous imagined speech studies [8, 19], and more generally, in most active BCI tasks [2]. Other than subject-specific performance of active mental tasks, this inconsistency could also, in part, be attributed to inter-individual variation in the shape and size of various brain regions. fMRI or similar imaging techniques could be used to confirm the brain regions interrogated at each of the 10-20 locations. Therefore, without the use of fMRI and structural data for each individual, it is not possible to assign a 10-20 location to a specific brain region and make a claim about the performance of a specific brain region [45].

In order to illustrate how [HbO] changed during a trial in channel 5, which provided the highest average and standard deviation of Fisher criterion and was approximately the closest channel to Wernicke's area [45], a graph illustrating [HbO] versus time, averaged over all trials of each participant is shown in figure 7. Individualized activation patterns were elicited in "yes", "no" and rest trials, which may be attributable to subject-specific performance of imagined speech. Unsurprisingly, the difference among the three trial types was generally more visually discernable in the data of participants with the highest classification accuracies.

### 4.3 Toward an asynchronous 2-class BCI

The control task in this study was an "unconstrained rest". In other words, participants were only asked to refrain from performing the other two imagined speech tasks during these trials. Hence, for the users who obtained a reliable 3-class accuracy, this synchronous BCI can be extended to a 2-class asynchronous BCI which can be activated by mentally repeating the phrase "yes" or "no". This asynchronous BCI can be used as a binary switch for an assistive device (i.e. with two activation modes) which the user can activate to call his/her caregiver, or to start a music player, for example. Using "yes" or "no" mental tasks to activate an assistive device would not be as intuitive as answering yes or no questions, but these tasks can be easier to perform than some of the common protocols of current fNIRS-BCIs (e.g. mental arithmetic) [17].

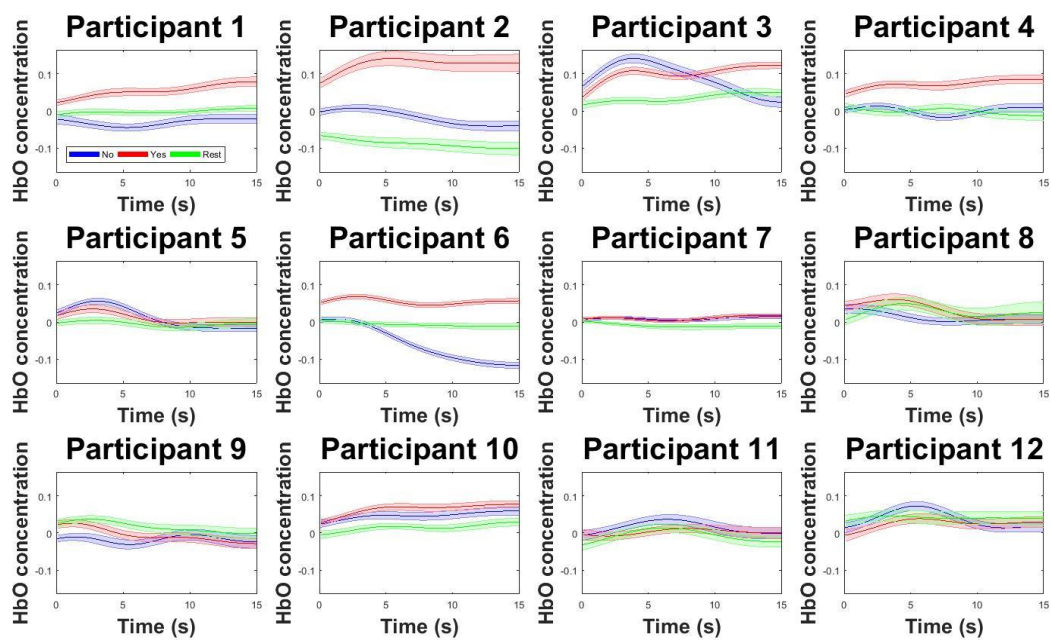

**Figure 7.** Averaged [HbO] responses recorded during the three trial types for channel 5. The shaded regions indicate the standard errors computed across all trials of the same class.

Depending on the application and preference of each user, the sensitivity and specificity levels for each activation task can be tuned. For example, if the task is of high importance, such as activating a call bell for assistance, the user may prefer a high sensitivity setting to err on the side of caution. On the other hand, if accidental activations are unwanted, such as switching on and off a music player, a higher specificity may be warranted.

*4.4 Limitations and future directions*

For future studies, the authors suggest the use of additional sessions, as the increased number of trials may enhance classifier performance. This study demonstrated that during the second session, the average accuracies during the last online block were significantly higher than those of the first online block, which is possibly due to the increased training data. Additional sessions would shed further insight on the achievable classifier performance and robustness. Also, prior to clinical translation, the findings herein must be replicated with individuals who present as locked-in.

Future research could also explore additional BCI-specific of additional BCI-specific intuitive commands, such as "left", "right", "stop" and "go" for navigation. Using words other than "yes" and "no" will also reveal if the classification results obtained herein were mainly due to users' intention to provide an affirmative versus a negative response, or due to the difference between covert articulation of "yes" and "no". Also, for future imagined speech BCI studies on able-bodied participants, an ultrasound system could be used to detect and discard trials which may contain possible motor confounds associated with subvocalization.

Finally, as each modality has been individually applied to the classification of imagined speech, a combination of EEG and fNIRS may exploit the advantages of each modality, potentially leading to improved BCI performance.

## 5. Conclusion

This study investigated an intuitive 3-class BCI based on imagined speech. Our findings suggest that fNIRS is a suitable modality for reliably differentiating affirmative and negative responses from unconstrained rest for selected BCI users. An average online classification accuracy of 64.1±20.6% was reached across all participants in the last three online blocks with nine participants exceeding the chance level ($p<0.001$). Task-related differences in the left temporal and left temporoparietal regions tended to provide discriminatory information. The proposed BCI could eventually empower individuals with severe disabilities with an intuitive means of interacting with their environment. To our knowledge, this is the first report of an online fNIRS 3-class classification of imagined speech.

## Acknowledgements

The authors would like to thank Dr. Frank Rudzicz, Dr. Stephen Strother and Dr. Silvia Orlandi for their guidance throughout this project. Special thanks also go to all the members of the PRISM Lab. This work was supported by the Natural Science and Engineering Research Council of Canada. Funding bodies had no involvement in the study, nor in the decision to publish.